\documentclass[%
 reprint,
 amsmath,amssymb,
 aps,
]{revtex4-2}

\usepackage{graphicx}
\usepackage{dcolumn}
\usepackage{bm}
\usepackage{amssymb}
\usepackage{epstopdf}
\newcommand{\scs}{\scriptscriptstyle}
\newcommand{\smallm}{\scs (\!-\!)}
\newcommand{\smallp}{\scs (\!+\!)}

\newcommand{\smallpm}{\scs (\!\pm\!)}

\graphicspath{{Fig/}}
\begin{document}

\preprint{APS/123-QED}


\title{Hybrid Tamm and quasi-BIC microcavity modes}

\author{D. S. Buzin}

\affiliation{Kirensky Institute of Physics, Federal Research Center KSC SB
RAS, Krasnoyarsk, 660036 Russia}
\affiliation{Siberian Federal University, Krasnoyarsk, 660041 Russia}%

\author{P. S. Pankin}

\email{pavel-s-pankin@iph.krasn.ru}

\affiliation{Kirensky Institute of Physics, Federal Research Center KSC SB
RAS, Krasnoyarsk, 660036 Russia}
\affiliation{Siberian Federal University, Krasnoyarsk, 660041 Russia}%

\author{G. A. Romanenko}

\affiliation{Krasnoyarsk Scientific Center, Siberian Branch, Russian Academy of Sciences, Krasnoyarsk, 660036 Russia}

\affiliation{Kirensky Institute of Physics, Federal Research Center KSC SB
RAS, Krasnoyarsk, 660036 Russia}

\affiliation{Siberian State University of Science and Technology, Krasnoyarsk, 660037 Russia}

\author{D. N. Maksimov}

\affiliation{Kirensky Institute of Physics, Federal Research Center KSC SB
RAS, Krasnoyarsk, 660036 Russia}
\affiliation{Siberian Federal University, Krasnoyarsk, 660041 Russia}%

\author{V. S. Sutormin}

\affiliation{Kirensky Institute of Physics, Federal Research Center KSC SB
RAS, Krasnoyarsk, 660036 Russia}
\affiliation{Siberian Federal University, Krasnoyarsk, 660041 Russia}%

\author{S. V. Nabol}

\affiliation{Kirensky Institute of Physics, Federal Research Center KSC SB
RAS, Krasnoyarsk, 660036 Russia}
\affiliation{Siberian Federal University, Krasnoyarsk, 660041 Russia}%

\author{F. V. Zelenov}

\affiliation{AO NPP Radiosvyaz, Krasnoyarsk, 660021 Russia}

\affiliation{Siberian State University of Science and Technology, Krasnoyarsk, 660037 Russia}

\author{A. N. Masyugin}

\affiliation{AO NPP Radiosvyaz, Krasnoyarsk, 660021 Russia}

\affiliation{Siberian State University of Science and Technology, Krasnoyarsk, 660037 Russia}

\author{M. N. Volochaev}

\affiliation{Kirensky Institute of Physics, Federal Research Center KSC SB
RAS, Krasnoyarsk, 660036 Russia}

\author{S. Ya. Vetrov}

\affiliation{Siberian Federal University, Krasnoyarsk, 660041 Russia}%

\affiliation{Kirensky Institute of Physics, Federal Research Center KSC SB
RAS, Krasnoyarsk, 660036 Russia}

\author{I. V. Timofeev}

\affiliation{Kirensky Institute of Physics, Federal Research Center KSC SB
RAS, Krasnoyarsk, 660036 Russia}

\affiliation{Siberian Federal University, Krasnoyarsk, 660041 Russia}%

\date{\today}

\begin{abstract}
The microcavity in the form of a liquid crystal defect layer embedded in a one-dimensional photonic crystal is considered. The microcavity mode has a tunable radiation decay rate in the vicinity of a bound state in the continuum. It is demonstrated that coupling between the microcavity mode and a Tamm plasmon polariton results in hybrid Tamm-microcavity modes with a tunable Q factor. The measured spectral features of hybrid modes are explained in the framework of the temporal coupled mode theory.
\end{abstract}

\maketitle

\newpage
\section{Introduction}

The Tamm plasmon polariton (TPP) is a surface electromagnetic wave localized at the interface between a metal and a one-dimensional photonic crystal (PhC) \cite{Vinogradov2006,kaliteevski2007tamm,sasin2008tamm}.
Unlike a surface plasmon polariton, TPP can be excited by optical radiation of any polarization without phase matching \cite{maier2007plasmonics}. Various optical devices such as lasers \cite{symonds2013confined}, optical switches \cite{zhang2010bistable, Afinogenov2019_TPP_Switch}, ideal absorbers \cite{gong2011perfect, lu2021nonreciprocal, bikbaev2023enhanced},
thermal emitters \cite{ChenKuoPing2017}, single photon sources \cite{Gazzano2012SPSTPP}, photodetectors \cite{huang2023wavelength}, and sensors \cite{zaky2023refractive, kumar2017self, maji2018blood} were proposed on the basis of the TPP.
The TPP can be coupled with other types of modes, with their simultaneous excitation in the system \cite{kar2023tamm, buzavaite2020influence, wu2023high}. In this case one speaks of hybrid TPP modes.
For example, when a PhC with a defect is coated with a metal layer, the TPP mode is coupled with the defect mode or, in other words, with the microcavity (MC) mode \cite{kaliteevski2009hybrid, Bruckner2011TPP_MC, Zhang2013TPP_MC, Zhang2015TPP_MC, toanen2020room}.
In our previous work \cite{pankin2021experimental} the control of the spectral position of hybrid TPP-MC modes in a PhC with a defect layer filled with a liquid crystal (LC) was experimentally demonstrated. The sensitivity of the LC to external factors makes it possible to efficiently control the MC modes as well as the TPP coupled with them.

The presence of an anisotropic layer in the PhC allows for optical bound states in the continuum (BICs) \cite{pankin2020one, liu2023bound}.
The BIC is a localized state that is embedded in the continuous frequency spectrum and is not coupled with the propagating modes \cite{HsuChiaWei2016, sadreev2021interference, Koshelev2023}. 
By changing the parameters of the system near the BIC, one can control the radiative decay rate of the resonant mode (quasi-BIC). This makes it possible to control the Q factor of the resonance and the amplitude of the resonant mode. Based on the BICs, numerous applications including lasers \cite{Kodigala17, wu2021quasi}, optical vortices \cite{huang2020ultrafast}, optical harmonics generation \cite{zograf2022high}, lidars \cite{huang2022tunable}, waveguides \cite{GomisBresco_Torner2017_BIC1D, Bezus2018_BIC}, perfect absorbers \cite{wang2023plasmonic}, sensors \cite{Romano2019_BICsensor} and cavities \cite{Rybin2017, jia2023bound} were proposed.

In the present work, we consider a microcavity in the form of a LC layer enclosed between two PhC mirrors. Such a layer acts as a MC which supports a quasi-BIC with a controlled radiative decay rate.
The presence of a gold (Au) layer at the top of the PhC results in simultaneous occurence of a TPP leading to hybrid TPP-MC modes.
In this paper, the spectral manifestation of the resonances induced by the hybrid modes are investigated theoretically and experimentally.   


\section{TCMT theory}

\subsection{MC mode}

\begin{figure}[h]
    \centering
    \includegraphics[width=0.5\textwidth]{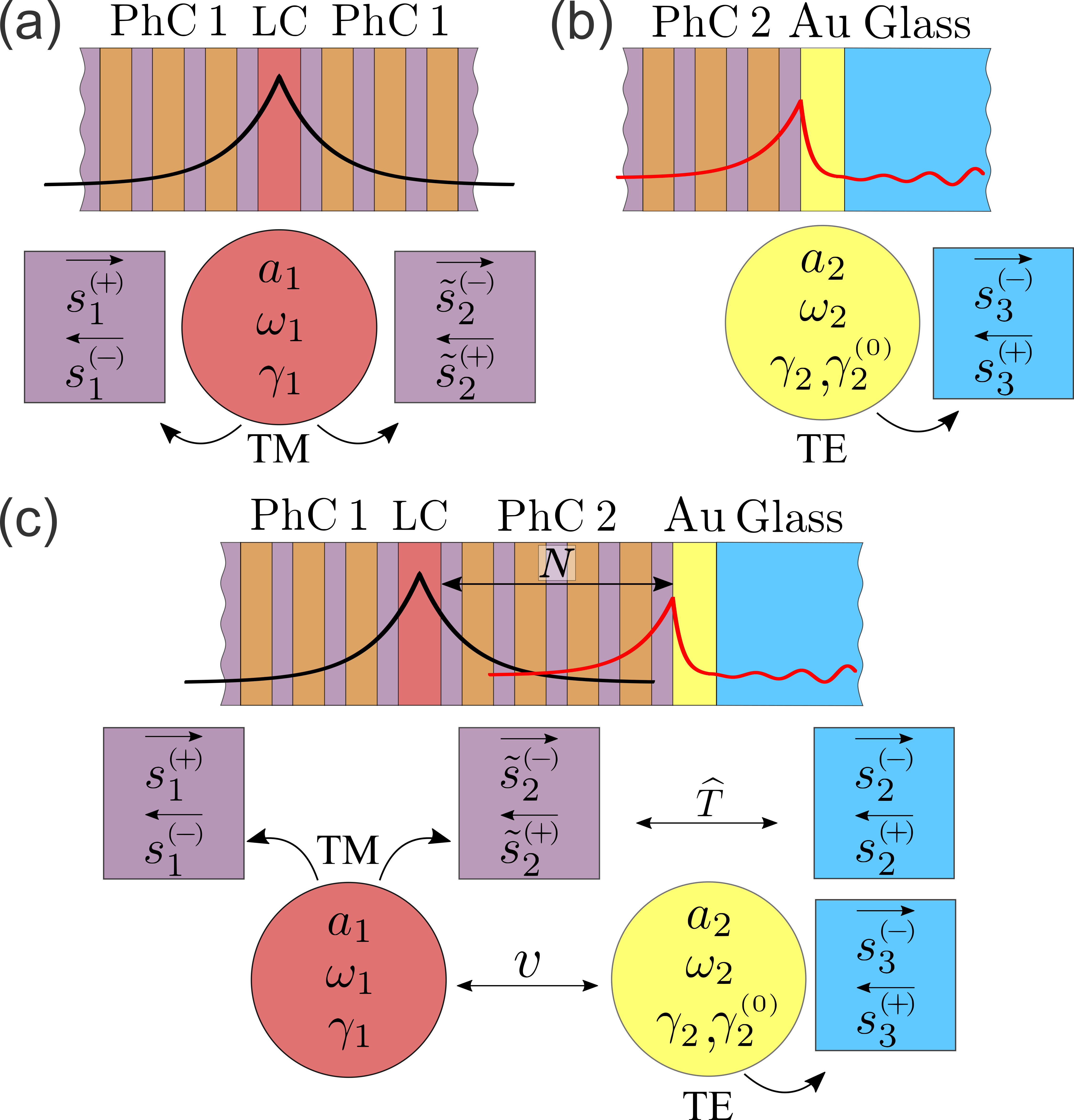}
    \caption{TCMT model for (a) MC-mode, (b) TPP-mode, (c) hybrid TPP-MC modes.} \label{fig1}
\end{figure}

The coupled-mode theory is a useful insturment for describing the features of the spectral behavior of photonic devices \cite{FanShanhui2003, gorkunov2020metasurfaces, Bykov2019_TCMTBIC, bulgakov2023optical}. Here within the framework of temporal coupled-mode theory (TCMT), we consider the resonant MC mode localized in the LC layer, which is coupled with scattering channels in semi-infinite PhCs, see Fig.~\ref{fig1}(a). 
When the Brewster  condition is satisfied \cite{Akhmanov1997bk}, the photonic band gap (PBG) vanishes for TM-polarized waves (TM waves). Therefore TM waves can propagate in the PhC, while TE-polarized waves (TE waves) undergo Bragg diffraction. In the layer of a planar-oriented LC, in the general case, ordinary ($o$-wave) and extraordinary ($e$-wave) waves can propagate and contribute to both, TM and TE waves.      
The amplitudes $s_{1}^{\smallp}$ and $\tilde{s}^{\smallp}_2$ correspond to incident TM waves, while $s_{1}^{\smallm}$ and $\tilde{s }^{\smallm}_2$ to reflected ones.
The equations for the amplitude $a_{1}$ of the MC-mode in the framework of TCMT \cite{FanShanhui2003} are written as 
\begin{align}\label{eqMC}
    & \frac{da_1}{dt}=-(i\omega_1+\gamma_1)a_1+\langle d^*|
    \left(
    \begin{array}{c}
         s_1^{\smallp} \\
         \tilde{s}_2^{\smallp}
    \end{array}
    \right),
\end{align}
\begin{align}\label{out}
    & \left(
\begin{array}{c}
s_{1}^{\smallm} \\
\tilde{s}_{2}^{\smallm} \\
\end{array}
\right)
=\widehat{B}
\left(
\begin{array}{c}
s_{1}^{\smallp} \\
\tilde{s}_{2}^{\smallp} \\
\end{array}
\right)+ a_1|d\rangle.
\end{align}
Here $\omega_{1}$ is the real part of the eigenfrequency of the MC mode, $\gamma_{1}$ is the imaginary part of the eigenfrequency of the MC mode, which is the radiative decay rate, $|d\rangle$ is the column vector of the coupling constants of the microcavity mode with TM wave channels. The radiation decay rate $\gamma_{1}$ depends on the orientation of the LC optical axis and vanishes when BIC is implemented \cite{krasnov2023voltage}. 
In the case of mirror symmetry with respect to the central plane, the direct process matrix $\widehat{B}$ can be written as 
\begin{equation}\label{direct}
\widehat{B}=
e^{i\psi}\left(
\begin{array}{cc}
\rho & \pm i\tau\\
\pm i\tau & \rho
\end{array}
\right),
\end{equation}
where $\rho$ and $\tau$ are the amplitudes of the complex reflection and transmission coefficients, respectively. The elements of the matrix $\widehat{B}$ are found as the scattering coefficients at the BIC frequency $\omega_{\mathrm{\scs{BIC}}}$.
According to \cite{maksimov2020optical}, the vector of coupling constants for a symmetric resonant mode is written as 
\begin{equation}
\label{d}
|d\rangle= \left(\begin{array}{l}d_1  \\ d_2 \end{array}\right) =
e^{i\frac{\psi}{2}}
\sqrt{\frac{\gamma }{2(1+\rho)}}
\left(
\begin{array}{l}
\pm\tau-i(1+\rho) \\
\pm\tau- i(1+\rho)
\end{array}
\right), 
\end{equation}
while for an antisymmetric mode takes the form
\begin{equation}
\label{dm}
|d\rangle= \left(\begin{array}{l}d_1  \\ d_2 \end{array}\right) =
e^{i\frac{\psi}{2}}
\sqrt{\frac{\gamma }{2(1+\rho)}}
\left(
\begin{array}{l}
\pm\tau+i(1+\rho) \\
\mp\tau- i(1+\rho)
\end{array}
\right).
\end{equation}
Assuming that harmonic waves $s^{\smallpm} \propto e^{-i\omega t}$ propagate in the channels, the simultaneous solution of Eq.~\eqref{eqMC} and Eq.~\eqref{out} gives the following expression for the scattering matrix $\widehat{S}$ near the BIC frequency $\omega_{\mathrm{\scs{BIC}}}$ of the MC mode
\begin{equation}\label{S}
\widehat{S}_0(\omega)=\widehat{B}+\frac{|d\rangle\langle d^*|}{i(\omega_1 - \omega)+\gamma_1}.
\end{equation}

\subsection{TPP mode}

Let us consider the case of a semi-infinite PhC covered with a semi-transparent layer of Au, see Fig.\ref{fig1}(b).
If Brewster's condition is satisfied, a TPP can be implemented only for TE waves.
The equations for the amplitude $a_{2}$ of the TPP mode within the framework of TCMT under the assumption of weak absorption in the metal are written as
\begin{align}\label{eqTPP}
     \frac{da_2}{dt}=-(i\omega_2+\gamma_2+\gamma_2^{\scs{(0)}})a_2+
    d_3s_3^{\smallp},
\end{align} 
\begin{align}\label{outTPP}
    {s}_{3}^{\smallm}=r_0{s}_{3}^{\smallp}+a_2d_3,
\end{align}
where $\gamma_2^{\scs{(0)}}$ is the rate of material loss in the Au layer, $r_0 = e^{i\eta}$ is the direct process reflection coefficient, $d_3 = \sqrt{2\gamma_2}e^{i(\eta - \pi)/2}$ is the TPP coupling constant with TE wave channel, ${s}_{3}^{\smallpm}$ are the amplitudes of TE waves in the glass substrate.
The joint solution of the Eq.~\eqref{eqTPP} and Eq.~\eqref{outTPP} gives the following expression for
the resonant reflection coefficient
\begin{equation}
\label{eta}
r(\omega) = r_0 +\frac{d_3^2}{i(\omega_2-\omega)+\gamma_2+\gamma_2^{\scs{(0)}}}.
\end{equation}

Further, we assume that the TPP frequency is unchanged and equal to the BIC frequency for the MC mode $\omega_2 = \omega_{\mathrm{\scs{BIC}}}$.
In the case of the critical coupling between the TPP and the incident wave $|r(\omega_{\mathrm{\scs{BIC}}})| = 0$, the phase of the direct process $\eta$ is determined by the following expression
\begin{equation}
\label{r}
\eta = arg[r(\omega_{\mathrm{\scs{BIC}}} + \gamma_2 + \gamma_2^{\scs{(0)}})] + \pi/4.
\end{equation}
If the critical coupling is not realized $|r(\omega_{\mathrm{\scs{BIC}}})| \ne 0$, then 
\[ \eta =
  \begin{cases}
    arg[r(\omega_{\mathrm{\scs{BIC}}})],       & \quad \gamma_2 > \gamma_2^{\scs{(0)}}\\
    arg[r(\omega_{\mathrm{\scs{BIC}}})] + \pi,  & \quad \gamma_2 < \gamma_2^{\scs{(0)}}.
  \end{cases}
\]

\subsection{Hybrid modes}

If both resonances are present in the system, the TCMT equation, namely Eq.~\eqref{eqMC} and Eq.~\eqref{eqTPP} can be combined into 
\begin{align}\label{eqHybrid}
    & \frac{da_1}{dt}=-(i\omega_1+\gamma_1)a_1 - iva_2+\langle d^*|
    \left(
    \begin{array}{c}
         s_1^{\smallp} \\
         \tilde{s}_2^{\smallp}
    \end{array}
    \right), \nonumber \\
    & \frac{da_2}{dt}=-(i\omega_2+\gamma_2+\gamma_2^{\scs{(0)}})a_2  - iva_1
    + d_3s_3^{\smallp}, 
\end{align}
where $v$ is the tunneling coupling constant between the resonators due to the overlap between the evanescent tails in PhC~2, see Fig.~\ref{fig1}(c) \cite{nabol2022fabry}.
After the time harmonic substitution Eq.\eqref{eqHybrid} for the amplitudes $a_1$ and $a_2$ can be written as
\begin{widetext}
\begin{equation}\label{two_modes}
      \left(
    \begin{array}{cc}
          i(\omega_1-\omega)+\gamma_1 & iv \\
         iv & i(\omega_2-\omega)+\gamma_2+\gamma_2^{\scs{(0)}} 
    \end{array}
    \right)
\left(
\begin{array}{c}
    a_1  \\
      a_2
\end{array}
\right) 
= 
  \left(
\begin{array}{c}
     d_1{s}_1^{\smallp} + d_2\tilde{s}_2^{\smallp} \\
     d_3s_3^{\smallp}
\end{array}
\right).   
\end{equation}

The ${s}_2^{\smallpm}$ TM wave amplitudes in the glass substrate are related to the $\tilde{s}_2^{\smallpm}$ amplitudes in PhC~2 by the transfer matrix $\widehat{T}(\omega)$ for the glass/Au/PhC structure
\begin{equation}\label{Tmatrix}
\left(
\begin{array}{c}
     \tilde{s}_2^{\smallp} \\
      \tilde{s}_2^{\smallm}
\end{array}
\right)
=\widehat{T}(\omega)
\left(
\begin{array}{c}
    {s}_2^{\smallp} \\
      {s}_2^{\smallm}
\end{array}
\right).
\end{equation}
Having expressed the unknown amplitudes $\tilde{s}_2^{\smallpm}$ in PhC~2 from Eq.~\eqref{Tmatrix} we can rewrite Eq.~\eqref{two_modes} in the following form
%
\begin{align}\label{two_modes+s}
& \left(
    \begin{array}{cc}
          i(\omega_1-\omega)+\gamma_1  - \frac{d_2^2T_{12}}{T_{22}-B_{22}T_{12}} & iv \\
         iv & i(\omega_2-\omega)+\gamma_2+\gamma_2^{\scs{(0)}} 
    \end{array}
    \right)
\left(
\begin{array}{c}
    a_1  \\
      a_2
\end{array}
\right) = 
&
\left(
\begin{array}{c}
     {s}^{\smallp}_1(d_1 + \frac{d_2B_{21}T_{12}}{T_{22}-B_{22}T_{12}}) + s^{\smallp}_2\frac{d_2}{T_{22} - B_{22}T_{12}} \\
     s_3^{\smallp}d_3
\end{array}
\right).   
\end{align}
%
By setting $s^{(+)}_{1,2,3}$ = 0, one can find the roots of the resulting equation, which are equal to the eigenfrequencies of the hybrid TPP-MC modes $\tilde{\omega}_{r1,2}$.
The joint solution of Eqs.~\eqref{out},~\eqref{outTPP},~\eqref{Tmatrix} and \eqref{two_modes+s} gives the final expression for the scattering matrix $\widehat{S}$ of the system relating the wave amplitudes in the scattering channels $| {s}^{\smallm}\rangle = \widehat{S} |{s}^{\smallp}\rangle$, see APPENDIX section 
\begin{equation}\label{Bigmatrix}
  \widehat{S}(\omega) = \left(\begin{array}{ccc}
   C_{11} + \Omega_{11}D_{1}^2  & C_{12} + \Omega_{11}D_{1}D_{2} & \Omega_{12}D_{1}d_3 \\
  C_{21} + \Omega_{11}D_{2}D_{1} & C_{22} + \Omega_{11}D_{2}^2 & \Omega_{12}D_{2}d_3\\
   \Omega_{21}d_3D_{1} & \Omega_{21}d_3D_{2} & r_0+\Omega_{22}d_3^2
    \end{array}
    \right).
\end{equation}
\end{widetext}

\section{Numerical results}
\begin{figure*}[ht]

\center{\includegraphics[scale=0.8]{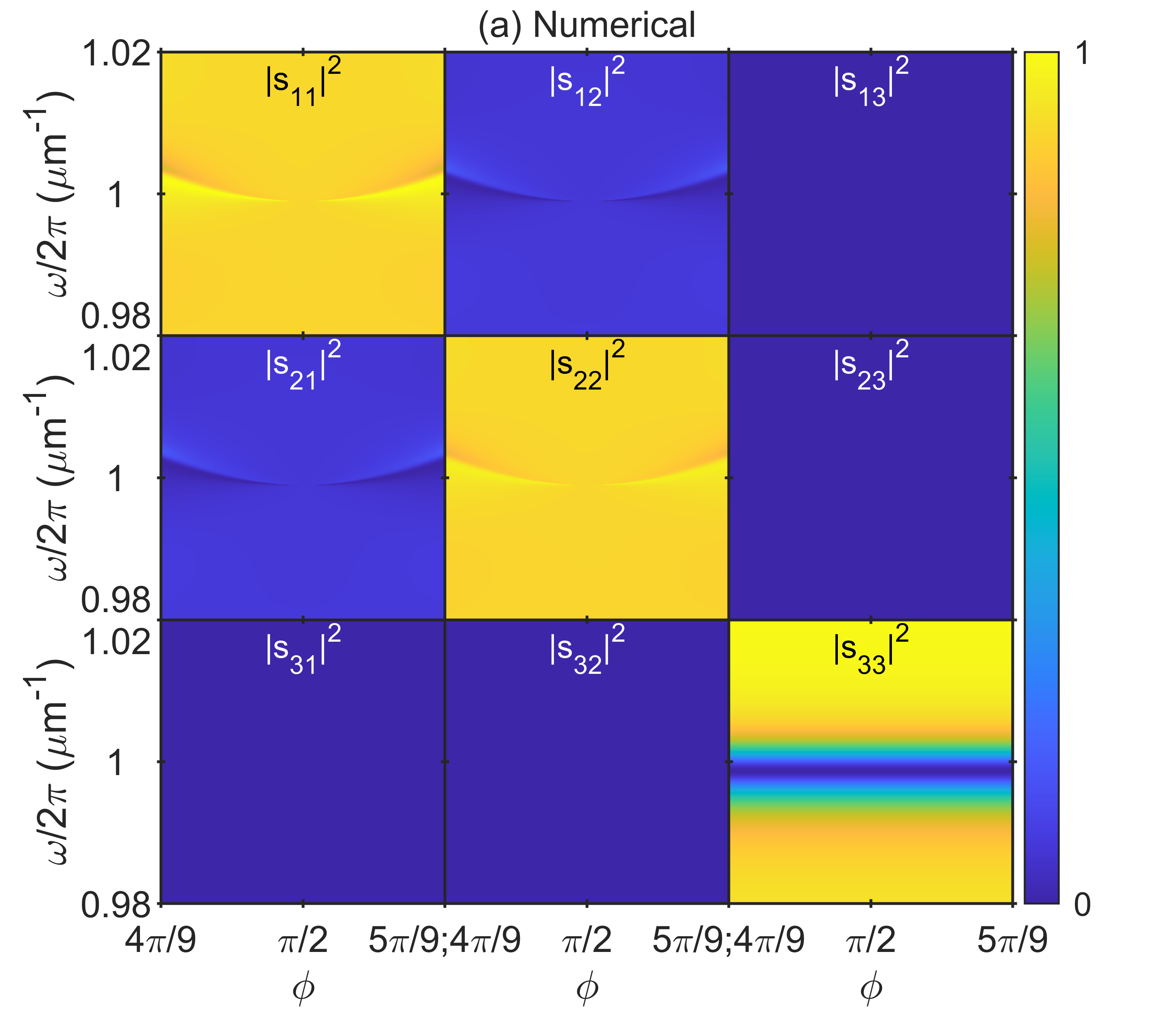}
\includegraphics[scale=0.8]{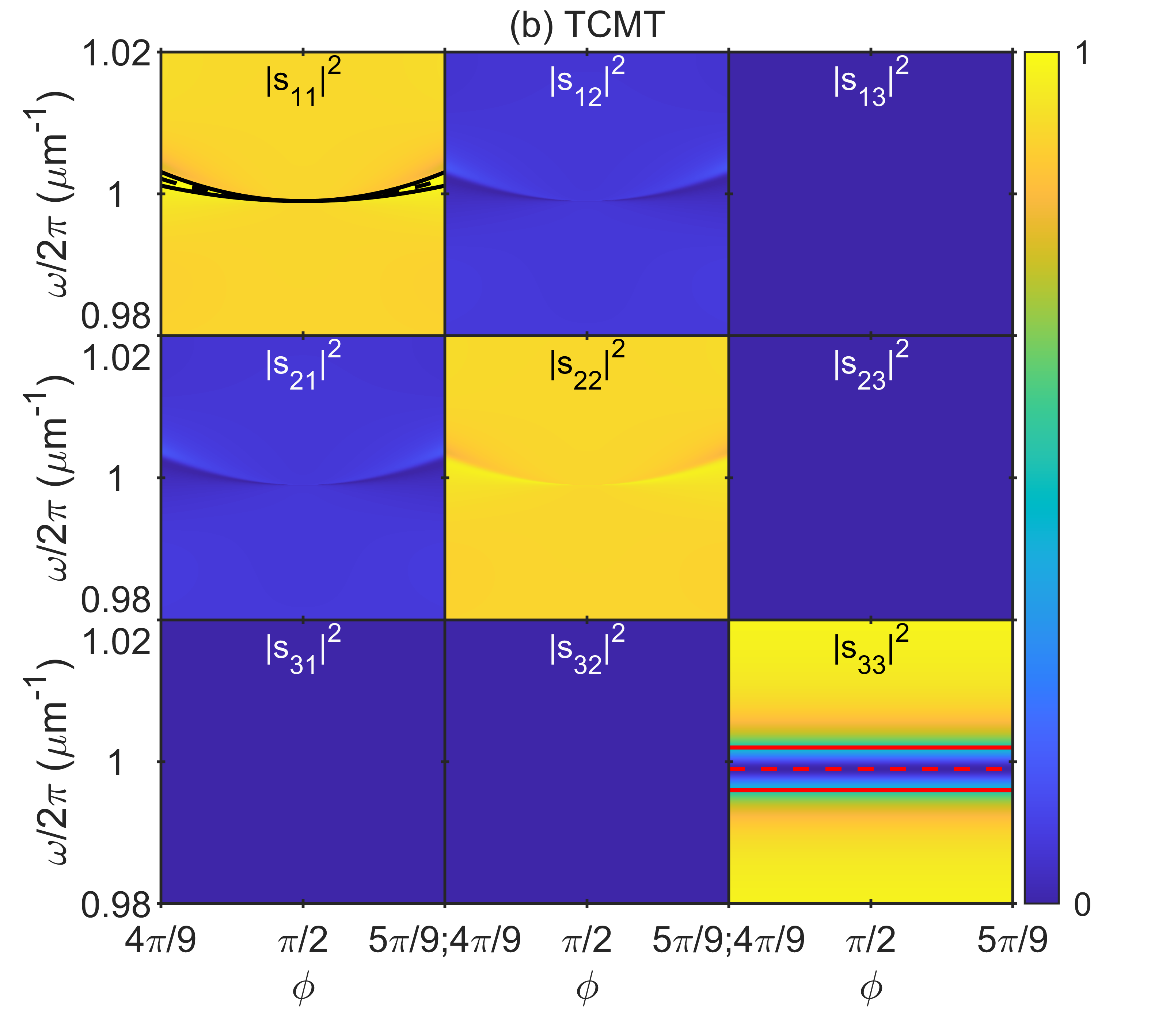}}
\caption{Scattering coefficients of the microcavity obtained (a) by the Berreman matrix method and (b) from the elements of the $\widehat{S}$-matrix Eq.~\eqref{Bigmatrix}. The angle of incidence $\theta_{B} = \arctan{(2/1.5)}$ from the glass substrate with the refractive index of 1.5. The refractive indices and thicknesses of quarter-wavelength PhC layers are $n_{1} = 2$, $h_{1} = 0.25/[2\cos{(\pi/2 -\theta_{B})]}$, $n_{2} = 1.5 $, $h_{2} = 0.25/[1.5\cos{(\theta_{B})]}$. Refractive indices and thickness of the LC layer are $n_{\parallel} = 1.7$, $n_{\perp} = 1.5$, $h_{LC} \approx 1.25\mu$m. The refractive index and thickness of the Au layer are $h_{Au} = 40$ nm, $n_{Au} \approx 0.11+6.47i$. The PhC~2 top layer thickness is $0.84h_1$. The number of periods in PhC~1 and PhC~2 is 30 plus one unpaired layer with refractive index $n_1 = 2$. Subplot (b) shows $\omega_{1}$, black dashed line; $\omega_{1} \pm \gamma_{1}$, black solid line; $\omega_{2}$, red dashed line; $\omega_{2} \pm (\gamma_{2} + \gamma_2^{\scs{(0)}})$, red solid line. The tunneling coupling constant $v = 0$.}
\label{TCMT_Ber_30per}
\end{figure*}

\begin{figure*}[ht]
\center{\includegraphics[scale=0.8]{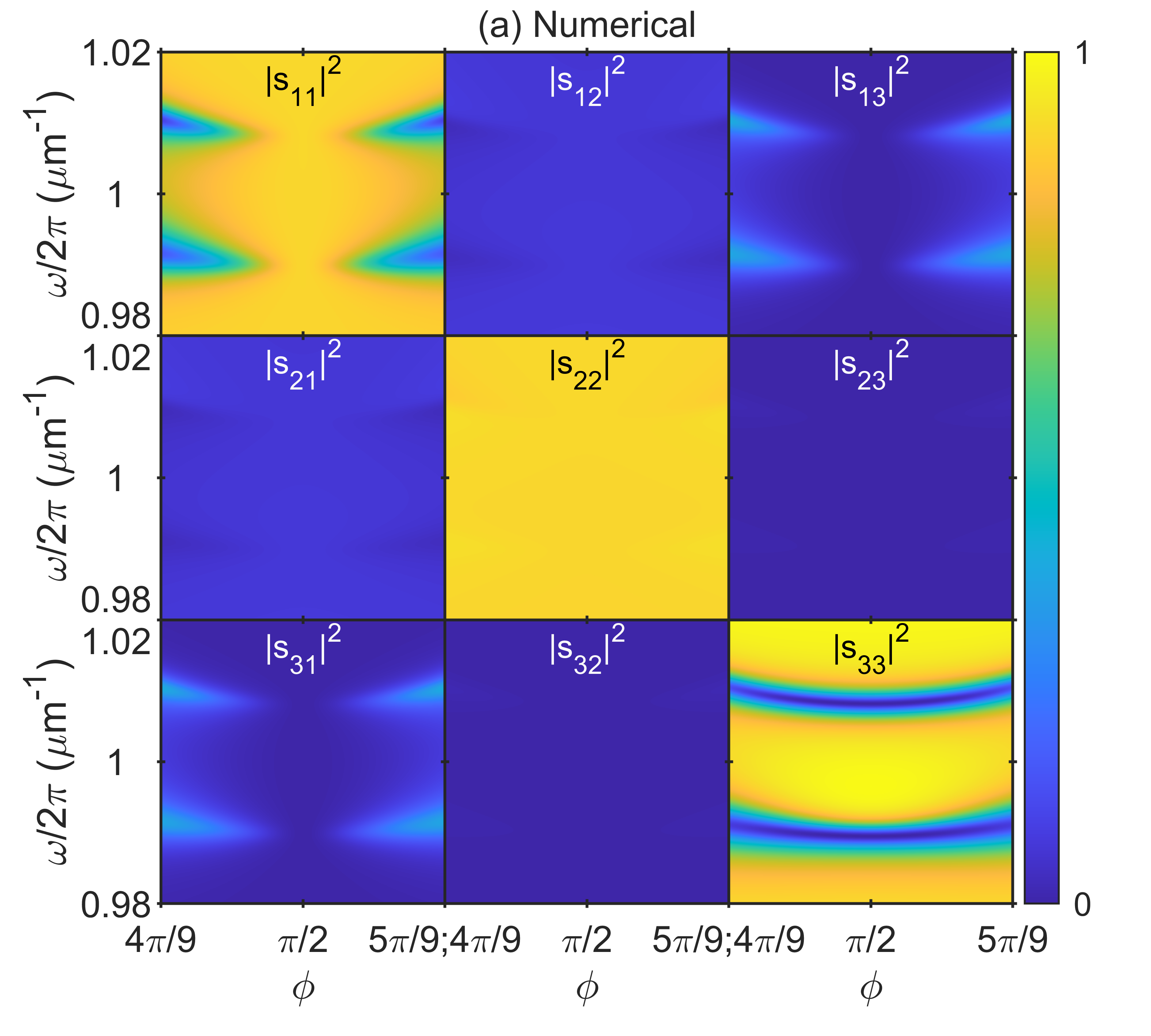}
\includegraphics[scale=0.8]{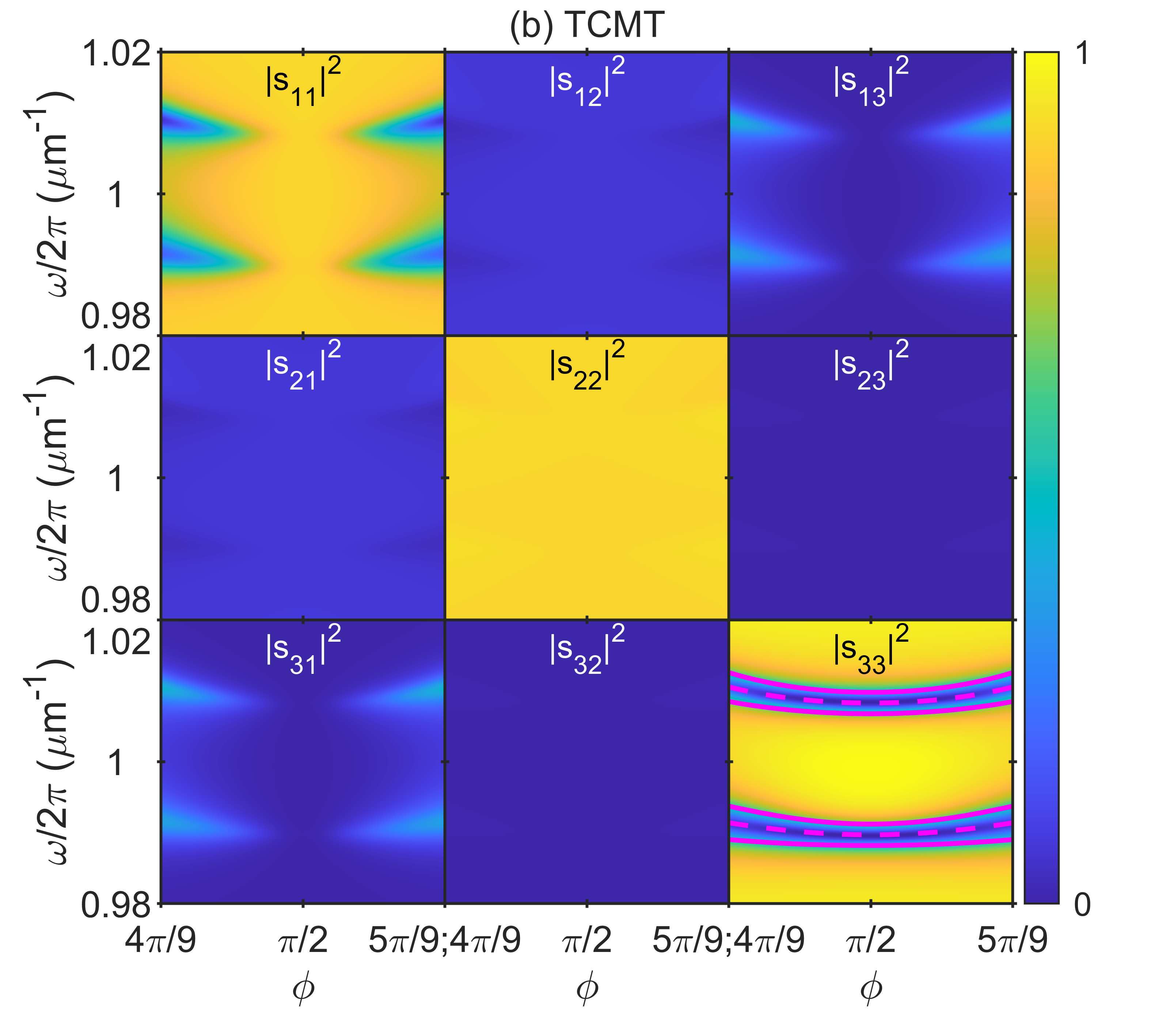}}
\caption{Scattering coefficients of the microcavity obtained (a) by the Berreman matrix method and (b) from the elements of the $\widehat{S}$-matrix Eq.~\eqref{Bigmatrix}.
The number of periods in PhC~1 and PhC~2 are 30 and 4, respectively, plus one unpaired layer with refractive index $n_1 = 2$. 
Subplot~(b) shows $Re(\tilde{\omega}_{r1,2})$, magenta dashed line; $Re(\tilde{\omega}_{r1,2}) \pm Im(\tilde{\omega}_{r1,2})$, magenta solid line. 
The tunneling coupling constant $v = 6.4\gamma_2$. 
The other parameters are the same as those in the caption to Fig.~\ref{TCMT_Ber_30per}.}
\label{TCMT_Ber_4per}
\end{figure*}

Figures \ref{TCMT_Ber_30per} and \ref{TCMT_Ber_4per} show the scattering coefficients of the microcavity obtained by the  Berreman transfer matrix method \cite{Berreman1972} and from the elements of the TCMT $\widehat{S}$-matrix Eq.~\eqref{Bigmatrix}.
The spectra were calculated depending on the azimuthal angle $\phi$ of the optical axis of the planar LC layer. The parameters of the TCMT model were found as follows.
The real $\omega_{1,2}$ and imaginary $\gamma_{1,2}$, $\gamma_2^{\scs{(0)}}$ parts of eigenfrequencies $\omega_{r1,2} = \omega_ {1,2}-i\gamma_{1,2}$ of the uncoupled resonant modes were found analytically by using the wave matching method with radiative boundary conditions, see Supplementary in \cite{pankin2020one}.
The elements of the matrix $\widehat{B}$ were obtained from the reflection and transmission amplitudes of the TM waves, calculated by the Berreman method at the BIC frequency $\omega_{\mathrm{\scs{BIC}}}$ in a structure without an Au layer.
The elements of the $\widehat{T}(\omega)$ matrix were calculated over the entire frequency range by the Berreman method in a structure without a LC layer for the TM waves. The phase $\eta$ was calculated at the BIC frequency $\omega_{\mathrm{\scs{BIC}}}$ for the TE waves.
The tunneling coupling constant $v$ was chosen by fitting with the numerical spectrum.
  
Figure \ref{TCMT_Ber_30per} corresponds to a large number of periods $N = 30$ in PhC~2.
It can be seen from the figure that the MC mode appears only in the spectra of the 1st and 2nd channels, corresponding to the TM waves $|s_{11}|^2$, $|s_{12}|^2$, $|s_ {21}|^2$, $|s_{22}|^2$.
The TPP appears only in the reflectance spectrum of the 3rd channel of TE waves $|s_{33}|^2$.
The spectra obtained within the framework of TCMT are consistent with the numerical ones at $v = 0$, which indicates that there is no overlap of the evanescent tails of the two resonant modes due to the large number of periods $N$ in PhC~2.
When $\phi = \pi/2$ vanishing of the resonant line of the MC mode is observed, indicating the realization of a symmetry-protected BIC in the absence of coupling between the MC and TPP modes \cite{pankin2020one}.

Figure \ref{TCMT_Ber_4per} corresponds to a small number of periods $N = 4$ in PhC~2.
It can be seen from the figure that the resonant lines exhibit an avoided crossing, which indicates the occurrence of a strong coupling between the TPP and MC modes.
This is also confirmed by the value of the coupling constant $v = 6.4\gamma_{2}$ at which the spectra obtained within the framework of TCMT agree with the numerical ones.
The $|s_{33}|^2$ spectrum is superimposed with the real parts of the eigenfrequencies of the hybrid modes $Re(\tilde{\omega}_{r1,2})$, as well as the values of $Re(\tilde{\omega}_ {r1,2}) \pm Im(\tilde{\omega}_{r1,2})$, see Fig.~\ref{TCMT_Ber_4per}(b). 
The positions of the eigenfrequencies are consistent with the positions of the resonances of the hybrid TPP-MC modes.
It is also seen from Fig.~\ref{TCMT_Ber_4per} that the coupling leads to the realization of resonant transmission between channels of different polarizations which is clearly seen in the spectra of $|s_{13}|^2$, and $|s_{31}|^2$. The manifestation of resonant transmission in the spectra of $|s_{23}|^2$, $|s_{32}|^2$ is less noticeable, but can be enhanced by adjusting the parameters of the system.
It can be seen from the figure that the resonant lines corresponding to the hybrid TPP-MC modes do not vanish in the $|s_{33}|^2$ spectrum. This indicates the impossibility of realizing a true BIC for hybrid modes in the presence of a coupling between the MC mode and the TPP mode, for which there are always radiation decay rate $\gamma_{2} \neq 0$. 

\section{Metal-dielectric microcavity}

\begin{figure}[h]

\includegraphics[scale=1.4]{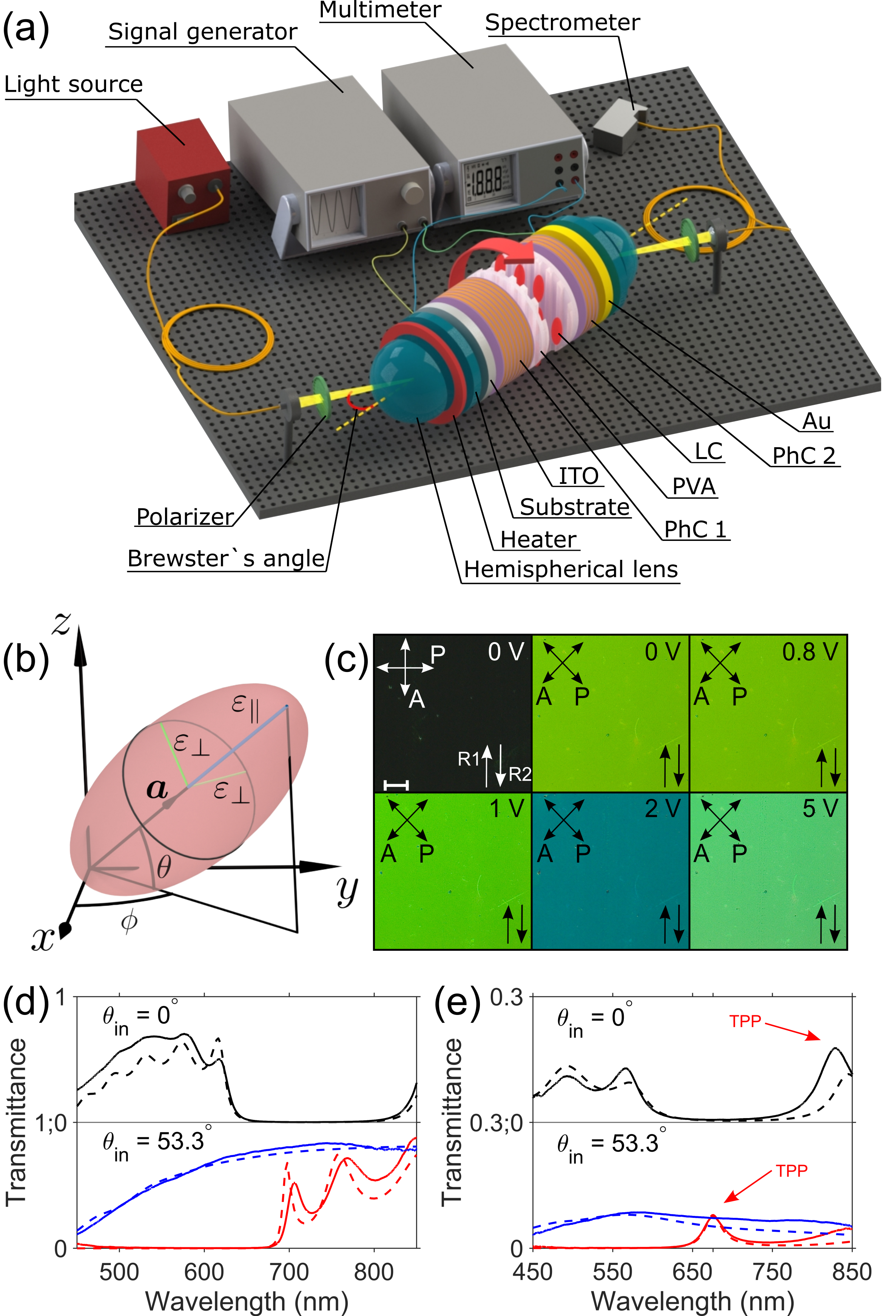}
\caption{\label{fig:ShemModel}
(a) Schematic representation of the experimental set-up. (b) Ellipsoid of the LC permittivity. (c) Photographs of the optical texture of the LC layer at different applied voltages U. The scale bar is $100 \; \mu$m in length.
(d,e) Measured, dashed line, and calculated, solid line, transmittance spectra of (d) PhC~1 and (e) PhC~2 at normal incidence, black line, and at Brewster's angle for TE, red line and TM waves,  blue line.}
\end{figure}

To implement hybrid TPP-MC modes, a microcavity with an anisotropic LC resonator layer was fabricated, see Fig.~\ref{fig:ShemModel}(a).
The microcavity consists of two PhCs - (PhC~1 and PhC~2), formed by periodically arranged layers of silicon nitride Si$_3$N$_4$ and silicon dioxide SiO$_2$ with refractive indices and thicknesses $n_{1} \approx 2.24$, $h_{1} \approx 79$~nm  and $n_{2} \approx 1.45$, $h_{2} \approx 135$~nm, respectively.
PhC~1 includes 8 periods and one unpaired Si$_3$N$_4$ layer.
PhC~2 including 4 periods and one unpaired Si$_3$N$_4$ layer was coated with a $h_{Au} \approx 46$ nm thick Au layer and 4 nm thick Ti adhesion layer to ensure the occurence of a TPP.
The PhC~2 top layer thickness is 88 nm.

The gap between PhCs formed by Teflon spacers was filled with nematic LC 4-pentyl-4'-cyanobiphenyl (5CB) by the capillary method.
The preferred direction of orientation of the LC molecules is determined by the unit vector $\bm{a}$, called the director.
The $\bm{a}$ vector determines the direction of the principal axis of the LC permittivity ellipsoid, which for nematics coincides with the LC optical axis, see Fig.~\ref{fig:ShemModel}(b).
The refractive index along and perpendicular to the director is determined through the components of the permittivity tensor $n_{\parallel} = \sqrt{\varepsilon_{\parallel}} \approx 1.73$, $n_{\perp} = \sqrt{\varepsilon_{\perp }} \approx  1.54$. For calculations the experimental data of the frequency-dependent refractive indices \cite{rodriguez2016self,luke2015broadband,rosenblatt2020nonmodal, palm2018dynamic, moerland2016subnanometer,Li2005_RI_E7,tkachenko2006nematic,Sefton19855CB_RI_Temp} were adjusted within 5$\%$ to provide agreement with the experimental spectra.

Both PhCs were covered with a 100 nm thick PVA layers.
After deposition, these layers were unidirectionally rubbed to ensure the planar orientation of the LC molecules.
Figure \ref{fig:ShemModel}(c) shows photographs of the optical texture of a LC layer using a ZEISS Axio Imager A1M optical microscope in a crossed polarizer scheme.
If the rubbing direction of the PVA layer coincides with the direction of the polarizer or analyzer, the optical texture is uniformly dark.
When the angle between the rubbing direction and the direction of the polarizer is $45^{\circ}$, the maximum transmission is observed.
This indicates a uniform planar orientation of the LC molecules in the resonator layer.
A 30 nm thick conducting layer of indium tin oxide (ITO) is placed between the substrate and PhC~1.
In PhC~2 the conducting layer is Au. When a 1 kHz AC voltage is applied to the conducting layers the color of the optical texture changes, which indicates a reorientation of the LC, see Fig.~\ref{fig:ShemModel}(c).

Figure \ref{fig:ShemModel}(a) shows a schematic representation of the experimental setup.
The white light from a Thorlabs OSL2 halogen lamp was launched through an optical fiber to a polarizer.
Linearly polarized light impinges on a microcavity with hemispherical glass lenses connected to the sample through immersion oil on both sides. Light is incident at angle $\theta_{in} = \theta_{B} = 53.3^{\circ}$ which ensures that the Brewster condition is satisfied at the boundaries of the PhC layers \cite{krasnov2023voltage}.
The light transmitted through the sample further passes through the analyzer and was collected by an optical fiber collimator connected to the OCEAN FX-UV-VIS spectrometer.
The change in the azimuthal angle $\phi$ of the LC optical axis was provided mechanically using the Thorlabs KPRM1E/M turntable.
The change in the polar angle $\theta$  of the LC optical axis was provided by a 1 kHz AC voltage applied to the conducting layers of ITO and Au from an Aktakom AWG-4150 generator.
Voltage control on the conductive layers was carried out using an Aktakom-4552 multimeter.
The sample temperature is set by a heater which was controlled by a thermistor.
All spectral measurements were automated and controlled by computer.           

Figures \ref{fig:ShemModel} (d,e) show the transmittance spectra of PhC~1 and PhC~2 at normal incidence $\theta_{in} = 0^{\circ}$ and at Brewster's angle $\theta_{in} = \theta_{B}$. The wide dip in the transmittance spectrum of PhC~1 corresponds to the PBG. When TE waves with $(0,E_y,0)$ components are incident at Brewster's angle, the PBG is blue shifted. At the same time, TM waves with components $(E_x,0,E_z)$ do not undergo Bragg reflection on the PhC periodic structure due to the Brewster effect, as evidenced by the absence of a dip in the transmittance spectrum.
Under normal incidence PhC~2 transmittance spectrum shows a resonant peak at 832 nm corresponding to the TPP, see Fig.~\ref{fig:ShemModel}(e).
When light is incident at the Brewster's angle, the resonant peak also is blue shifted and is observed only in the spectrum of TE waves at a wavelength of $\lambda_{\mathrm{\scs{TPP}}} \approx 670$ nm.

\section{Experimental results}

\begin{figure}[h]

\includegraphics[width = 1\linewidth]{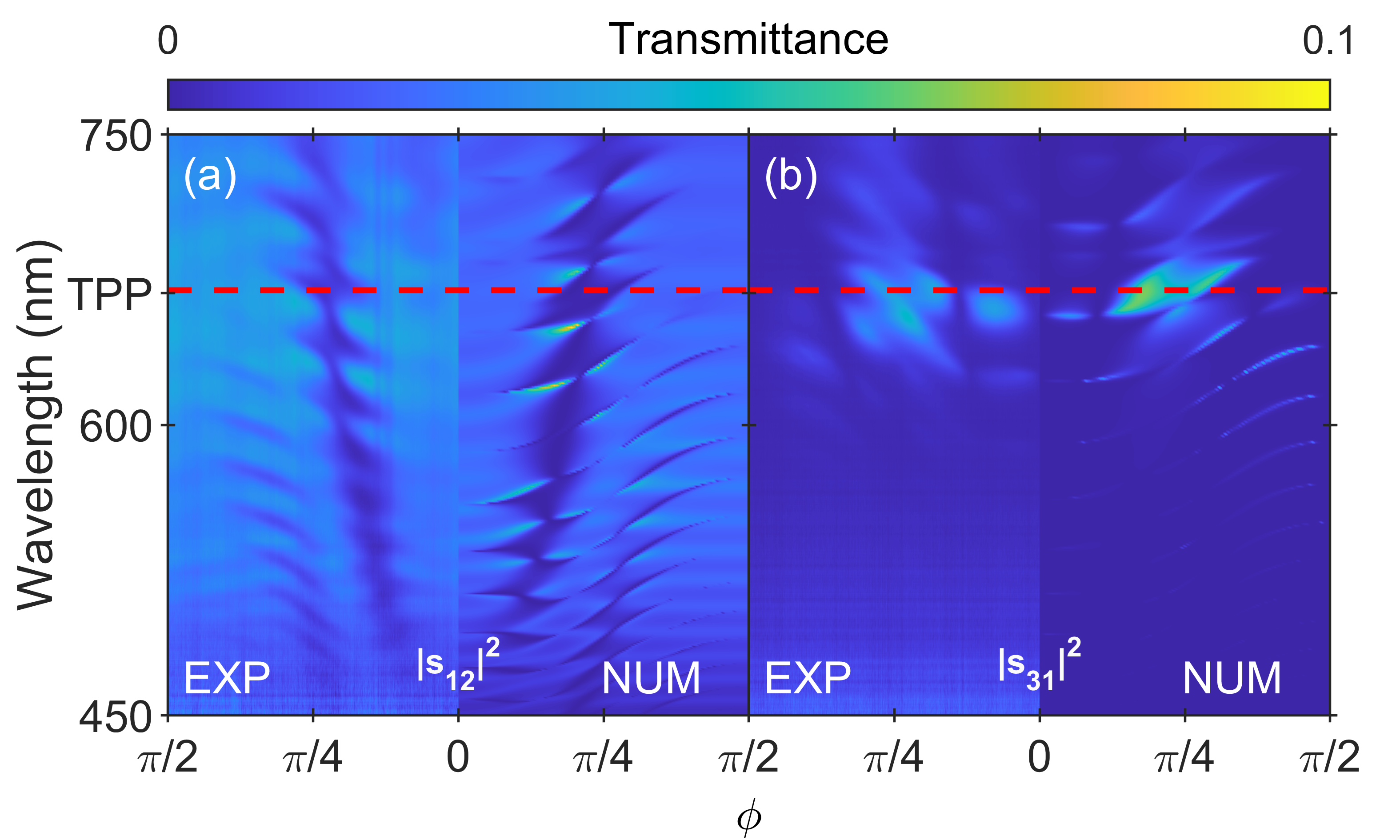}
\caption{\label{fig:meas} Measured and calculated dependences of transmittance spectra $|s_{12}|^{2}$ (a) and $|s_{31}|^{2}$ (b) on the azimuthal angle $\phi$. The red dashed line shows the spectral position of the TPP $\lambda_{\scs{TPP}}$. The LC layer thickness is $h_{LC} \approx 5.4 \; \mu$m.}
\end{figure}

Figure \ref{fig:meas}(a) shows the measured and calculated transmittance spectra $|s_{12}|^{2}$ of the microcavity for TM waves as a function of the azimuthal angle $\phi$ at $\theta = 0$.
The spectra show multiple resonant lines corresponding to the MC modes, which undergo a red shift when the angle $\phi$ increases from $0$ to $\pi/2$.
Many vanishing resonant lines are also observed in the spectra, which indicates realization of the BICs.
Far from the spectral position of TPP $\lambda_{\scs{TPP}}$, in the vicinity of the vanishing point of an individual resonant line of the MC mode, the single-resonance approximation is applicable, for which the $\widehat{S}_0$-matrix Eq.~\eqref{S} is derived.
Within the framework of TCMT the vanishing of spectral lines is explained by vanishing of the coupling constants $d_{1,2}$.
It can be seen from Eq.~\eqref{d} and Eq.~\eqref{dm} that the coupling constants $d_{1,2} \propto \sqrt{\gamma_1}$ vanish in the absence of radiation losses $\gamma_1 = 0$.
The latter are determined by the Poynting vector of TM waves of the resonant mode $\gamma_1 \propto |E_x|^2$ at the LC/PVA boundary \cite{Timofeev2018_BIC}.

There are two possibilities for satisfying the condition $E_{x} = 0$ on the LC/PVA boundary.
In cases when the  LC optical axis angle is $\phi = 0,\pi/2$, no mixing of TM and TE waves occurs in the LC layer.
Maxwell's equations are separable, and only TE waves contribute to the resonant mode \cite{pankin2020one}. This case corresponds to BICs protected by symmetry.
This also explains the red shift of the resonant lines, since the change of angle $\phi$ from $0$ to $\pi/2$ leads to the variation of the LC refractive index for TE waves from $n_{\perp}$ to $n_{\parallel }$.
In general, when $\phi\neq 0$, $\theta \neq 0$ the $o$-wave $\bm{E}_{o}$ and the $e$-wave $\bm{E}_{e}$ amplitudes in the LC layer have all Cartesian components $\bm{E}_{o,e} = (E_{o,ex}, E_{o,ey}, E_{o,ez})$, which contribute in both TM- and TE waves \cite{Ignatovich2012, krasnov2023voltage}. 
For certain non-trivial values of the angle $\phi$, the condition $E_x = E_{ox} + E_{ex} = 0$ is satisfied on the LC/PVA boundary. This case corresponds to Friedrich-Wintgen type BICs, also called accidental BICs or parametric BICs \cite{Koshelev2023}.

Figure \ref{fig:meas}(b) shows the measured and calculated transmittance spectra $|s_{31}|^{2}$ of the  microcavity corresponding to TM waves incident on PhC~1 and TE waves passing through PhC~2.
In contrast to the results shown in Fig.~\ref{TCMT_Ber_4per}, obtained in the approximation of TPP coupling with only one MC mode, TPP resonantly couples with several neighboring MC modes at once \cite{pankin2017tunable}.
Far from the TPP position $\lambda_{\scs{TPP}}$ there are no resonances of the MC modes in the $|s_{31}|^{2}$ spectrum. This confirms the resonant transmission due to the tunneling coupling.
The results obtained indicate the occurrence of hybrid TPP-MC modes.

\begin{figure}[ht]
\includegraphics[width = 1\linewidth]{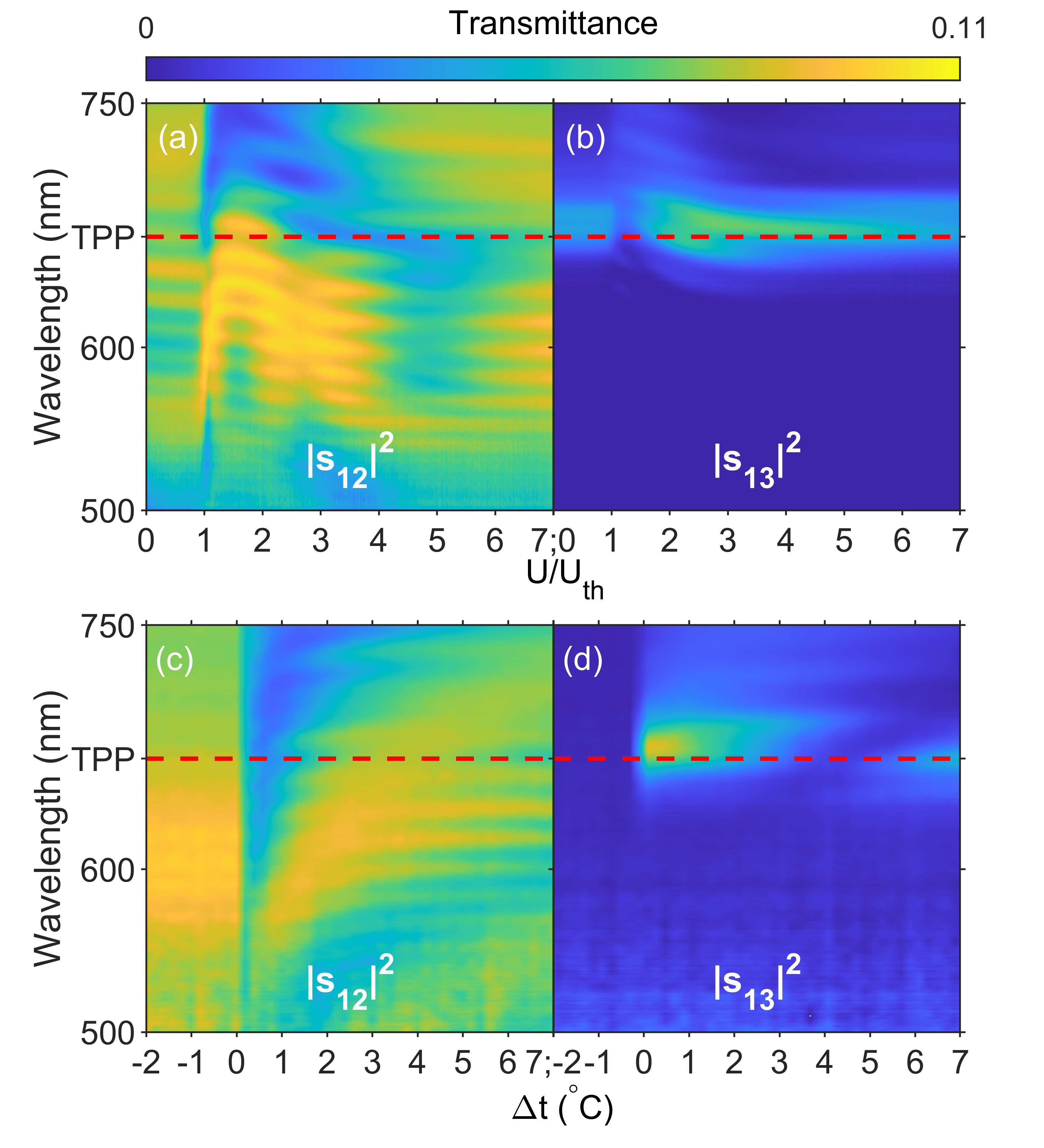}
\caption{\label{fig:measVoltTemp} The measured dependences of transmittance spectra of $|s_{12}|^{2}$ (a) and $|s_{13}|^{2}$ (b) on the applied voltage U. U$_{\text{th}} \approx 1$V is the threshold voltage for LC reorientation. The azimuthal angle $\phi = \pi/4$ (a), $\pi/3$ (b). The measured dependences of transmittance spectra of $|s_{12}|^{2}$ (c) and $|s_{13}|^{2}$ (d) on the difference $\Delta t = t_{0}-t$ between the temperature $t$ of the LC layer and the  temperature of the LC/isotropic liquid phase transition $t_{0} \approx 34.4^{\circ}$C. The azimuthal angle $\phi = \pi/4$. The red dashed line shows the spectral position of the TPP $\lambda_{\scs{TPP}}$. The LC layer thickness is $h_{LC} \approx 7.9 \; \mu$m.}
\end{figure}

Due to the sensitivity of the LC to external factors, it is possible to dynamically tune the spectra of the microcavity \cite{Ozaki2003LCdef,  Arkhipkin2011, Huang2016a, Pankin2021APL, krasnov2023voltage}.
Figure \ref{fig:measVoltTemp}(a,b) shows the transmittance spectra of the microcavity as a function of the voltage applied to the LC layer for fixed values of the azimuthal angle $\phi$.
At voltage U $=$ U$_{\text{th}}$, the spectra show an abrupt change in the position of the resonant lines, see Fig.~\ref{fig:measVoltTemp}(a).
This phenomenon is connected with a beginning of LC reorientation caused by the Frederiks effect \cite{Blinov2010bk}.
With a further increase in U, the LC molecules align in the direction of the external field applied along the $z$ axis, which leads to a change in the angle $\theta$, see Fig.~\ref{fig:ShemModel}(b).
In the limiting case of large U, the polar angle $\theta = \pi/2$, and the LC refractive index for the resonant mode has the smallest value $n_{\perp}$. This leads to a blue shift of the resonant lines of the MC modes.
The vanishing of the resonant lines is explained by the vanishing of the coupling constants $d_{1,2}$ as described above, when the condition $E_{x} = 0$ is satisfied on the LC/PVA boundary \cite{krasnov2023voltage}.
The $|s_{13}|^{2}$ spectrum shows resonant transmission in the $\lambda_{\scs{TPP}}$ region, see Fig.~\ref{fig:measVoltTemp}(b).
It can be seen from the spectrum that the MC mode coupled with the TPP in the vicinity of U $=$ U$_{\text{th}}$ moves away from the TPP position $\lambda_{\scs{TPP}}$ towards short wavelengths, and when the U increases the adjacent MC mode is coupled with the TPP.

Another dynamic control method is heating the LC layer \cite{Arkhipkin2008, wu2021quasi, Pankin2021APL}. Figure \ref{fig:measVoltTemp} (c,d) shows the transmittance spectra of the microcavity as a function of the temperature of the LC layer for fixed values of the azimuthal angle $\phi$.
The shift and vanishing of the resonant lines of the MC modes occurs due to the change in  $\varepsilon_{\perp}$ and $\varepsilon_{\parallel}$ with temperature, see Fig.~\ref{fig:measVoltTemp}(c) \cite{Sefton19855CB_RI_Temp, tkachenko2006nematic, Li2005_RI_E7}.
For certain temperatures, the phase shift of the $o$- and $e$-waves, determined by the values $\varepsilon_{\perp}$ and $\varepsilon_{\parallel}$, ensures that the condition $E_x = E_{ox} + E_{ex} = 0$ is satisfied at the LC/PVA boundary, which leads to  vanishing coupling constants $d_{1,2}$.
At the same time, the resonances of the hybrid TPP-MC modes also disappear in the $|s_{13}|^{2}$ spectrum, see Fig.~\ref{fig:measVoltTemp}(d).

\section{Conclusion}

In the present work, hybrid modes formed by coupling between a Tamm plasmon polariton and a microcavity mode were considered.
In the framework of the temporal coupled mode theory, the scattering problem in the spectral vicinity of a bound state in the continuum is solved.
The spectra obtained within the framework of the theoretical model are consistent with the spectra calculated by the exact Berreman matrix method.
The coupling between the TPP and MC modes manifests itself spectrally in the form of an avoided crossing of resonant lines and the resonant transmission from the TM polarization to the TE polarization channels.
For experimental verification of the results, the transmittance spectra of a microcavity with a resonant liquid crystal layer were measured.
The measured spectra are in qualitative agreement with the theoretical model.
In the transmittance spectra of the microcavity, multiple vanishing resonant lines of the microcavity modes were observed, indicating the realization of bound states in the continuum.
Resonant transmission of the TM into the TE polarization was observed near to the Tamm plasmon polariton spectral position.
Due to the sensitivity of the liquid crystal to external factors, dynamic control of the position and width of the spectral lines of hybrid modes was realized by mechanically rotating the sample, applying voltage to the liquid crystal layer, and heating the sample.
The results obtained can be used in the design of energy efficient photonics devices with a tunable Q factor.

\textbf{Acknowledgments.} This work was supported by the Russian Science Foundation under grant no 22-22-00687.
The authors would like to express their special gratitude to the Krasnoyarsk Regional Center for Collective Use of the Federal Research Center “Krasnoyarsk Scientific Center, Siberian Branch of the Russian Academy of Sciences” for providing equipment within this project.

\textbf{Disclosures.} The authors declare no conflicts of interest.

\textbf{Data Availability Statement.} The data that support the findings of this study are available from the corresponding author, P.S.P., upon reasonable request.

\section{Appendix}

The elements of $\widehat{S}$-matrix Eq.~\eqref{Bigmatrix} are:
\begin{align}
C_{11} = B_{11} + B_{12}^2T_{12}/(T_{22}-B_{22}T_{12}), \nonumber \\ 
C_{12} = C_{21} =  B_{12}/(T_{22}-B_{22}T_{12}), \nonumber \\ 
C_{22} = -(T_{21} - B_{22}T_{11} )/(T_{22}-B_{22}T_{12}),  \nonumber \\ 
D_{1} = d_1 + d_2B_{12}T_{12}/(T_{22}-B_{22}T_{12}), \nonumber \\
D_{2} = d_2/(T_{22}-B_{22}T_{12}), \nonumber \\
\widehat{\Omega} = \left(
    \begin{array}{cc}
          i(\omega_1-\omega)+\gamma_1  - \frac{d_2^2T_{12}}{T_{22}-B_{22}T_{12}} & iv \\
         iv & i(\omega_2-\omega)+\gamma_2+\gamma_2^{\scs{(0)}} 
    \end{array}
    \right)^{-1}. \nonumber
\end{align}

\bibliography{apssamp.bbl}

\end{document}